\documentstyle[prb,aps,ifthen,twocolumn]{revtex}
\input epsf

\def\NOTE#1 {}

\newcommand{\lsim}{{\lesssim}}  %%% replace this by the better symbol
\newcommand{\Estr}{{E^{\rm str}}}
\newcommand{\Eliq}{{E^{\rm liq}}}
\newcommand{\Esa}{{E^{\rm sa}}}
\newcommand{\HH}{{\cal H}}
\newcommand{\up}{{\uparrow}}
\newcommand{\dn}{{\downarrow}}
\newcommand{\mustar}{{\mu^*}}
\newcommand{\muLC}{{\mu^{\rm LC}}}

\newcommand{\ttilde}{{\tilde{t}}}
\def\dagg {{^\dagger}}
\def \PRLhead #1{{\it #1 --}}

\begin{document}

%%%% RESTORE THIS FOR TWOCOLUMN
\twocolumn[\hsize\textwidth\columnwidth\hsize\csname@twocolumnfalse%
\endcsname

\draft

\title{Spinless fermions and charged stripes
at the strong-coupling limit}
\author{C.~L.~Henley$^{a,b}$ and N.-G.~Zhang$^b$}
\address{$^a$Institute for Theoretical Physics, University of California, 
Santa Barbara, CA 93106}
\address{$^b$
Laboratory of Atomic and Solid State Physics, 
Cornell University, Ithaca, NY, 14853-2501}
\maketitle

\begin{abstract} 
Spinless fermions on a lattice with nearest-neighbor repulsion
serve as a toy version Hubbard model, and have
a symmetry-broken even/odd superlattice at half-filling.
At infinite repulsion, doped holes
form charged stripes which are antiphase walls
(as noted by Mila in 1994). 
%%% A single wall is an exactly soluble model, 
Exact-diagonalization data for systems up to 
36 sites around 1/4 filling, and also for one or two holes added to
a stripe of length up to 12, 
indicate stability of the stripe-array state
against phase separation.
%%% into the half-filled and liquid phases.
In the {\it boson} version of the model, 
the same behavior can be stabilized 
by addition of a four-fermi term.
\end{abstract}

%%% PACS numbers: 
%%% 71.10.Fd Lattice fermion models, 
%%% 71.10.Pm Fermions in reduced dimensions, 
%%% 05.30.Jp Boson systems, 
%%% 74.20.Mn Nonconventional mechanisms of superconductivity]}

%%%% RESTORE THIS FOR TWOCOLUMN
]

\narrowtext

%%% Introduction: Repulsive, spinless fermions in $d=2$}

Forty years of study have not yet 
produced a complete understanding of the phase diagram of the
Hubbard model, the simplest nontrivial paradigm of interacting 
spinfull fermions.\cite{Fa99}
%%%% Thus, it may be valuable to systematically study
The {\it spinless} lattice fermion model~\cite{Ko67} is a 
%%% strictly 
simpler and more tractable analog which retains many Hubbard-model properties, 
much as the Ising model 
stands in for the $n$-component magnet in critical phenomena:
understanding of the spinless model may provide fresh viewpoints of
the Hubbard model, or new tests of known methods.
Spinless models also arise naturally 
for ferromagnetic materials 
in which one of the spin-split bands is completely 
full or completely empty, 
such as magnetite \cite{Cu71} or ``half-metallic'' manganites~\cite{halfmetal}.

Our aim is to promote the systematic study of 
this model's phase diagram, which is almost untouched 
in the literature~\cite{FN-halffilled,Mu95,uhrig}. 
As a beginning, 
this  paper argues that, in the strong-coupling limit, 
the spinless model possesses a phase with the quantum-fluctuating, 
hole-rich antiphase domain walls known as ``stripes.''
Such stripes are an active topic in the 
Hubbard or $t-J$ 
models,~\cite{emery,Pre94,zaanen,Na97,Es98,white,pryadko,newzaanen}
particularly since stripes were observed in cuprates \cite{Tr94} and
seem related to incommensurate correlations
found in high-temperature superconductors.

Let us take a square lattice model with Hamiltonian
   \begin{equation}
   \HH = -t \sum _{\langle ij \rangle } (c_i\dagg c_j + c_j\dagg c_i)
   + V \sum _{\langle ij \rangle }  \hat{n}_i \hat{n}_j   
   \label{eq:Hamspinless} 
   \end{equation}
Here $c_i\dagg$ and $c_i$ are creation/annihilation
operators on site $i$, 
$\hat{n}_i\equiv c\dagg_i c_i$, and ``$\langle ... \rangle $'' 
counts each nearest-neighbor pair once. 
In most places we will consider hard-core bosons in parallel with
fermions \cite{bosons-XXZ}.
From here on, we take 
$V/|t|=\infty$ so  neighboring particles are simply forbidden, and
$t$ is the only energy scale.
(This constraint amounts to adopting a 
hardcore radius just over 1 lattice constant.)

\PRLhead{Phase diagram as function of $n$}
%%% $n \ll 1/2$ 
Consider first the dilute limit, $n  \lsim 0.15$.
When $V \to \infty$, the Hartree-Fock approximation
gives absurd results; 
in reality the renormalized interaction of two particles
is of order $t$ and a Bose or Fermi liquid is expected, 
as when the on-site repulsion $U \to \infty$ 
in the Hubbard model~\cite{t-matrix}.
%%% The moderate $n$ regime is discussed later in this paper.

At the other extreme, the dense limit ($n=1/2$, half-filling)
admits only the two microstates
with the $\sqrt 2 \times \sqrt 2$ checkerboard pattern, 
called the ``CDW'' (``charge-density wave'') order~ \cite{FN-halffilled}.
%%% with fermions preferentially on the even or odd lattices. 
An Ising symmetry breaking between even/odd lattices is exhibited, 
the spinless model's cartoon of the Heisenberg antiferromagnetic 
order at half-filling in the large-$U$ Hubbard model.
%%% or $t$-$J$ model.

\begin{figure}
  \centerline{\epsfxsize=2.4in\epsfbox{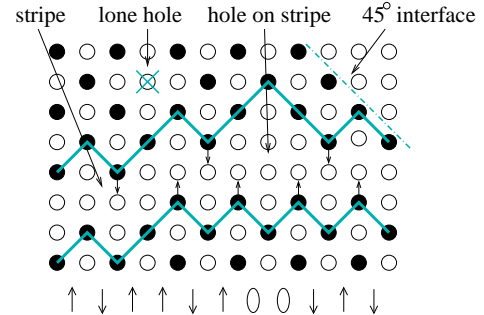}}
  \caption{
Hard-core particles (filled circles) must form a 
checkerboard 
at half-filling in the $V=\infty$ limit.
No hopping is possible by the lone hole ($\times$ mark) or 
$45^\circ$ interface (dot-dash line), 
but hopping is easy (arrows) along the stripe. 
The stripe's edges are shown by shaded lines 
and its spin-chain representation is below.
%%% Steps up or down map to up or down spins.
One extra hole is shown on the stripe.}
\label{fig:config}
\end{figure}

\PRLhead{Stripes in a hard-core model}
Now consider light hole doping, $1/2- n \ll 1$.
An isolated hole is immobile and 
gains no hopping energy in the CDW state
(see Fig.~\ref{fig:config}).
As Mila \cite{mila} observed, 
a droplet including $\ge$3 holes can
fluctuate but is still
confined to a circumscribed
rectangle with edges along the $45^\circ$ directions, 
since a particle is prevented from hopping away from a CDW domain surface 
oriented along $\{ 11 \}$ (Fig.~\ref{fig:config}).

The natural way to dope holes is a ``stripe'', an
antiphase domain wall with charge 1/2 hole per unit length.
This permits hops (arrows in Fig.~\ref{fig:config})
which implement stripe fluctuations.~\cite{mila}
A single stripe's path 
can be parametrized as a unique function 
$y(x)$ (hopping never generates overhangs).
Then $y(x+1)-y(x) = \pm 1$, and the steps up or down 
can be represented by a string
of corresponding $\up$ and $\dn$ arrows.
%%% (See bottom of Fig.~\ref{fig:config}.)`
A particle hop has the effect $\up\dn \to \dn\up$ and
so the Hamiltonian of a single stripe
maps exactly to the spin-1/2 XX chain~\cite{mila}
with exchange $J_\perp = -t$ for the X and Y spin components.
That model is exactly soluble by
a well-known mapping, whereby  the up (down) spins map 
to {\it noninteracting} spinless fermions  (empty sites), 
respectively, in one dimension. 
%%% (moving in a vacuum $\dn ... \dn$).
It follows that the energy of the (coarse-grained) stripe is 
     \begin{equation}
         \int dx [\sigma_0 + {\scriptsize \frac{1}{2}} K (dy/dx)^2 + \ldots],
     \label{eq:stripeE}
     \end{equation}
where $\sigma_0 = -(2/\pi) t$ and the stripe stiffness $K = (\pi/2) t$.
%%% and the path has been coarse-grained so that $dy/dx$ can be defined 
As usual, the sound velocity of the stripe's capillary waves
is the Fermi velocity $v=2 t / \hbar $ of the 1D noninteracting fermions.
Knowing $K$ and $v$ allows us to compute the fluctuations of the
Fourier mode at each wavevector $q$, as for any harmonic string: 
$\langle |y_q|^2 \rangle = \hbar v / 2K |q|$. 
The general result is that one such ``Gaussian'' 
quantum-fluctuating stripe has divergent fluctuations,
     \begin{equation}
         \langle |y(x)-y(0)|^2\rangle 
         = (v/2\pi K) \ln |x| + {\rm const}, \quad\hbox{as $x\to\infty$}
     \label{eq:stripefluc}
     \end{equation}
where $v/2\pi K = 2/\pi^2$ in this case.

The thermodynamic phase at $1/2-n \ll 1$ could then be
an array of stripes~\cite{FN-walls}
all parallel (on average) to either
the $\hat x$ or $\hat y$ axis.
%%% separated by $l=1/2\delta$ and incommensurate
%%% to a first approximation; 
They have only a contact interaction, so the array's long-range order
depends on stripe collisions, which 
surely exist since isolated stripes have divergent fluctuations
(eq.~(\ref{eq:stripefluc})). 
%%% OMIT twofold discrete symmetry breaking (stripe orientation). 
%%%%%%%%%%%%%%%%%%%%%%

Prior analytic work~\cite{Mu95} suggested that
spinless fermions in $d=2$, when doped away from half-filling,
develop an incommensurate ordering wavevector 
slightly off from $(\pi, \pi)$).
%%%%%%%%
%%%%%%%%%% 
Expanding around the $d=\infty$ limit,\cite{uhrig} 
small doping led to {\it coexistence} between the half-filled CDW
and a slightly incommensurate state
(but not at $V/t \to \infty$).
We conjecture these incommensurate phases, in $d=2$, 
consist of stripe arrays.
%%%%%%%%%%%%%%%%
%%%%%%%%%%%%
In the hard-core boson model near half-filling, 
in a regime $0 < V/t < \infty$, 
the uniform CDW phase is asserted to phase-separate 
upon doping~\cite{batrouni}.
The dense coexisting state is just as plausible, {\it a priori}, 
to be a stripe array as the phase-separated state 
that was assumed.~\cite{batrouni,batrounipc}.

\PRLhead{Stability estimates}
The key question is whether (or when) the stripe-array is stable,
compared to a phase-separated state in which the CDW state
and the dilute (=hole-rich) liquid coexist.
In the case of the Hubbard model, 
it was argued that doping invariably
leads to phase separation~\cite{visscher}
%%% (coexisting antiferromagnetic insulator and hole-rich metal), 
except when it is ``frustrated'' \cite{emery}
by the long-range Coulomb force.  
Contrarily, it was argued that holes in
fluctuating stripes may gain more kinetic energy than they 
would in a phase-separated state~\cite{Pre94}.

\begin{figure}
  \centerline{\epsfxsize=2.5in\epsfbox{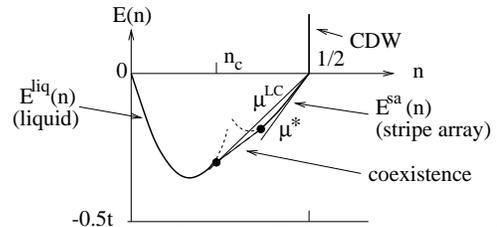}}
  \caption{
Schematic of energy $E(n)$ as a function of filling $n$, 
assuming $\muLC < \mustar$.
%%% according to ~(\ref{eq:Endilute}) and (\ref{eq:EnCDW}).
Dotted segments are metastable states.
%%%  thin line is coexistence between states marked by dots
Thin lines are drawn with slopes $\mustar$ and $\muLC$.}
  \label{fig:En}
\end{figure}

To decide the issue of coexistence, one first plots the 
energy per site $\Eliq(n)$  and $\Esa(n)$ for
the low-density liquid and the stripe-array, respectively, 
%%%% as a function of density $n$, 
which should look like Fig.~\ref{fig:En} for either fermions or 
hardcore bosons.  We have
%%%%%%%%%%%%%%%%%%%%%%%%%%%%%%
   \begin{equation}
         \Eliq(n) = (-4t)n + A_2 n^2 + A_3 n^3 + \ldots
   \label{eq:Endilute}
   \end{equation}
%%%%%%%%%%%%%%%%%%%%%%%%%%%%%%
where the leading coefficient 
%%% in (\ref{eq:Endilute}) is exact:
is the bottom of the single-particle band.~\cite{FN-dilboson}
With increasing density, the energy 
$\Eliq(n)$ turns upwards
and becomes small around $n=0.3$ as the
hopping becomes ``jammed''
(neighbor sites become forbidden due to other adjacent particles). 
The matrix elements contribute with the same sign 
in the boson ground state but can't in the fermion case, 
so $\Eliq_{\rm fermions} > \Eliq_{\rm bosons}$.

On the other hand, in the conjectured stripe array 
near half filling,
%%%%%%%%%%%%%%%%%%%%%%%%%%%%%%
   \begin{equation}
         \Esa(n)\approx 
%%% (-\scriptsize{\frac{4}{\pi}} 
[\sigma_0 + \phi(d)]  (1-2n)
    \label{eq:EnCDW}
   \end{equation}
%%%%%%%%%%%%%%%%%%%%%%%%%%%%%%
where 
$\sigma_0$ is the energy per unit $x$-length from 
(\ref{eq:stripeE});
the second factor is the length of stripe per unit area.
%%% (recall a stripe has $1/2$ hole per unit length), 
The mean stripe separation is $d\equiv (1-2n)^{-1}$, and
%%% The leading coefficient in (\ref{eq:EnCDW}) is
%%% $\sigma_0 = -2t/\pi $  from (\ref{eq:stripeE}). 
$\phi(d)$ parametrizes the energy cost per unit length
from collisions of adjacent stripes,~\cite{newzaanen} 
%%%%% with an average separation $d$, 
so $\phi(d) \to 0$ as $d \to \infty$.
Thus, the chemical potential in the limit of separated stripes
is $\mustar \equiv d\Esa(n)/dn|_{n=1/2} = (4/\pi)t = 1.273t$. 
We emphasize that the form
and the {\it leading} coefficients in (\ref{eq:Endilute}) 
and (\ref{eq:EnCDW}) are the same for fermions and hardcore bosons.
[Indeed, the fermion and boson models are identical in the single-stripe
sector, since the $V=\infty$ constraint prevents any
permutations;
this identity extends to the single stripe with one
extra hole, in which case only even permutations are 
accessible~\cite{stripehole}.]

There are three necessary conditions for the stability
of the stripe array: we have strong evidence for each 
of them, from exact diagonalizations of systems with 
20 to 72 sites.~\cite{NG}
These are far too small for direct observation
of a fluctuating stripe array, or of the coexistence of the liquid
and dense phases; yet they are large enough to yield some of
the parameters which the phase diagram can be calculated from.
From here on we use units $t\equiv 1$.

The first stability condition is that stripes repel, i.e. $\phi(d)>0$; 
%%%%%%%%%%%%
%%%%%%%%%%%%
stripe attraction would suggest instability to a domain of liquid phase, 
which is scarcely distinguishable from a bundle of self-bound stripes.
In both the fermion and boson cases
we diagonalized $L\times L'$ systems, 
%%% with $L$ and $L'$ both even and $LL'\leq 40$, 
for $L=4$, and $L'=6,8,10$, as well as $L=L'=6$, 
%%% and $L= 6$, $L' \leq 8 (?)$, 
doped with $L$ holes so that two stripes run in
the short direction, and thus $d=L'/2$. 
Define a stripe interaction per unit length
$\phi_{\rm eff} (L'/2) \equiv (E-2\Estr(L))/2L$, where
$\Estr(L)$ is the energy of one isolated stripe of
finite (even) length $L$. ($\Estr(L)$ is calculated exactly using
the 1D spinless-fermion representation.)
%%%%%%%%%%%%
%%%%%%%%%%%%
%%%%%%%%%%%
Indeed, $\phi_{\rm eff}(d)$ was positive and (for $L=4$)
decreasing with $d$.

Next, a stripe can contain extra holes,~\cite{Na97}
which move as quasiparticles with an excitation gap $\Delta$. 
The second stability condition is 
%%%%%%%%%%%%%%%%%%%%%%%%%%%%%%
   \begin{equation}
   \Delta > -\mustar
   \label{eq:E0ineq}
   \end{equation}
%%%%%%%%%%%%%%%%%%%%%%%%%%%%%%
If not, further doping would add holes to existing
stripes rather than form new ones,
again suggesting a tendency to form phase-separated droplets.
From diagonalizations we measured
the excitation energy of one added hole,~\cite{FN-stripetostripe}
$\Delta(L,L')$, with even $L$ and odd $L'$ in the range
$[4,8]$  for fermions, or $[4,12]$ for bosons;
as mentioned above, this is strictly 
independent of statistics~\cite{stripehole}.
From this we extrapolated, first $L'\to \infty$ using
$\Delta(L,L') = \Delta(L) + A_\Delta (L)e^{-L'/l(L)}$
and then $L\to\infty$ using $\Delta(L)=\Delta + B_\Delta/L$.
We found $\Delta = -0.65(5) $, which comfortably satisfies (\ref{eq:E0ineq}).

We also analyzed the two-hole energy $\Delta_2 (L)$, 
for $L= 4,6,8,10$ only;  with two holes, the $L$ 
dependence is more like $1/2^L$ than $1/L$. 
Extrapolating to $L=\infty$, 
yielded $\Delta_2 =-1.42(7)$ for bosons and $-1.44(9)$ for fermions.
Note $\Delta_2 - 2\Delta \approx -0.1$, i.e. hole binding
is insignificant when $L\le 10$;
we think it is a real effect in a large system, 
since holes on a stripe can be collected into
a (hole-free) {\it vertical} segment of the stripe. 
(Since the stripe's $90^\circ$ kinks cost energy, 
an array of parallel stripes will still be the
thermodynamic phase in a large system.)
%%%%%%%%%%%%

The third condition is the crucial one: 
as shown in Fig.~\ref{fig:En}, the dense phase coexisting with 
the liquid must not be the CDW, which would preempt a stripe-array phase.
%%% The phase separation should be liquid and stripe array.
That is, 
%%% Then the liquid-CDW line lies above $\Esa(n)$, 
%%%%%%%%%%%%%%%%%%%%%%%%%%%%%%
   \begin{equation}
       \mustar > \muLC  . 
   \label{eq:mustar}
   \end{equation}
%%%%%%%%%%%%%%%%%%%%%%%%%%%%%%
where $\muLC$ is the slope of the 
trial liquid-CDW tie-line 
tangent to $\Eliq(n)$ and passing through $(n=1/2, E=0)$. 

To test (\ref{eq:mustar}), the equation of state $\Eliq(n)$ is
required.  We exactly diagonalized
all rectangular lattices with $L, L'\ge 4$ and $LL'=20$ to 36 sites,
and with occupation in the range $0.20 \leq  n < 0.3$, and fitted
the results to  (\ref{eq:Endilute}). 
We obtained $(A_2, A_3) \simeq (9.45\pm 0.6,5\pm 2)$ for bosons and 
$(11.25\pm 0.6, -1\pm 2)$ for fermions.
This implies
$\muLC_{\rm bosons} = 1.33(2)$ and 
$\muLC_{\rm fermions} = 1.25(2)$. 
(Here the errors are estimated by varying 
the subset of data used for the fit.) 
For either bosons or fermions, coexistence with the CDW
would occur at $n_c\approx 0.24$.
%%%%%%%%%%%%

Hence, stripes are unstable in the boson case and (very likely)
stable in the fermion case, but close enough to the boundary
in either case that the balance can be tipped 
either way by the small perturbation $\ttilde_c$, discussed later.

\begin{figure}
  \centerline{\epsfxsize=1.9in\epsfbox{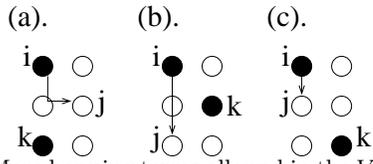}}
  \caption{
More hopping terms allowed in the $V=\infty$ Hilbert space, from
site $i$ to site $j$ when site $k$ is occupied (Eq.~(\ref{eq:newhops}).
These terms might modify the phase diagram (see text).}
%%% Configurations of $i,j,k$ in various
%%% versions of the occupation-dependent hopping term (\ref{eq:newhops}).}
\label{fig:newhops}
\end{figure}

\PRLhead{Exotic states?}
Like the Hubbard model, the spinless fermion model may be extended 
by adding other hopping terms to the Hamiltonian, which might
stabilize additional phases. 
Many of these terms have the form 
   \begin{equation}
         -\ttilde_x (c\dagg_j c_i + c\dagg_i c_j) \hat{n}_k, 
   \label{eq:newhops}
   \end{equation}
where $(i,j,k)$ are three different sites arranged as in 
Fig.~\ref{fig:newhops}, and $x$ is ``$a$'', ``$b$'', or ``$c$''
for the hops shown in the corresponding parts of
Fig.~\ref{fig:newhops}. 
For example, when $V$ is large but finite, hops are possible to a 
neighbor's neighbor with $\ttilde_a=\ttilde_b= -t^2/V$  as in
Fig.~\ref{fig:newhops}(a) and (b), analogous to similar
terms of order $t^2/U$ when the $t$-$J$ model is derived from
the Hubbard model.
This spinless analog of the $t$-$J$ model, in
which virtual states with neighbor pairs are projected out,
%%%  via second-order perturbation, or a canonical transformation.]
will be the natural starting point to study phenomena at large 
(but not infinite) $V$, {\it e.g.} the mobility of lone holes. 

In the fermion model, 
one could artificially take $\ttilde_b \gg t$ (analogous to
$J \gg t$ in the $t$-$J$ model);
then the term Fig.~\ref{fig:newhops} (b) naturally favors superconductivity.
Namely, fermions form
tightly bound $p$-wave pairs, 
separated by $\sqrt 2$;
these composite bosons hop with bandwidth $8 \ttilde_b$, 
and Bose-condense in the usual fashion.
Thus $\ttilde_b$
is analogous to  negative $U$ in the Hubbard model, in that 
superconductivity is put in ``by hand''.
But it is a  plausible speculation that,  
in the highly correlated liquid  at $n \approx 0.25$, 
BCS superconductivity appears even with $\ttilde_b \ll t$.

Finally, consider the hopping $\ttilde_c$ 
of Fig.~\ref{fig:newhops} (c) which 
%%% (since $i$ and $j$ are nearest neighbors)  
just modifies the amplitude of already possible hops.
This tends to stabilize (destabilize) stripes 
according to whether $\ttilde_c$ has the same
(opposite) sign as $t$, 
since every allowed hop in a stripe is surrounded by particles 
on all four possible ``$k$'' sites.
(Compare Fig.~\ref{fig:newhops}(c) with Fig.~\ref{fig:config}). 
Hence the stripe energy $\sigma_0$ and $\mustar$ get multiplied by a
factor $(1+4 \ttilde_c/t)$. 
On the other hand, assuming that each 
``$k$'' site is about $1/3$ occupied in the liquid at $n=0.25$, 
it follows that $\Eliq(n)$ is multiplied by
about $(1+4 \ttilde_c/3 t)$. 
If so, the critical perturbation where $\muLC=\mustar$
(so the stripe phase appears or vanishes)
is only $\ttilde_c/t \approx 0.02$ for bosons or
$-0.007$ for fermions, using
our values of $\muLC$ quoted above. 

%%%%%%%%%%%%%%%%%%%%%%%%%%%%%%%%%%%%%%%%%%%%%%%

\PRLhead{Discussion}
To establish the occurrence of stripes in the $V=\infty$ system, 
we addressed, by exact diagonalizations, 
(i) stripe-stripe interactions;
(ii) the energy of a single hole, 
as well as hole-hole interactions, 
on a stripe; and
(iii) the medium density liquid regime.  
(iv) hole-hole interactions on a stripe 
%%%%%%%%%%%%%%%%%%%%%%%%%%%%%%%
%%%%%%%%%%%%%%%%%%%%%%%%%%%%%%%
The enormous reduction of Hilbert space due to
the nearest-neighbor exclusion (at $V=\infty$), as well as 
the lack of spin, permits numerical explorations
at system sizes much larger than would be possible in
the Hubbard model 
-- vital not just for studying stripes, but 
any microscopically inhomogenous states. 
Monte Carlo simulation of the 
stripe phase is straightforward for the hardcore boson case.~\cite{batrounipc}
Boson results are valid for fermions too, 
when the stripe separation $d$ is large {\it and}
the density of ``extra holes'' on each stripe is low,
since 
particles do not exchange in this limit~\cite{stripehole}.
As for the fermion case, the new ``meron-cluster'' Monte Carlo algorithm 
cancels the sign problem for a limited class of models
{\it including spinless fermions}, but not the basic Hubbard model
\cite{wiese}.

The ultimate aim of microscopic simulations should be to extract
macroscopic parameters, {\it e.g.} the stripe stiffness $K$ or
the stripe contact repulsion. This is more straightforward 
than in the spin-full (Hubbard or $t$-$J$) case, where the
inter-stripe domains contain gapless spin-wave excitations~\cite{pryadko}.
These parameters may be input to analytic explorations
of the interesting anisotropic conductivity
of the quantum-fluctuating stripe array~\cite{Noda99}, 
also simpler in the spinless case.

More broadly, it is a challenge to test for the 
exotic phases we mentioned 
in connection with Fig.~\ref{fig:newhops}.
In the medium-density regime $n\approx 0.2$,  strong correlations
of {\it some} sort are essential to minimize the hopping energy.
These might be prosaic, {\it e.g.} a $\sqrt 5 \times \sqrt 5$
superlattice, but the following possibilities are realizable, 
in principle, even in a spinless model:
(i) orbital magnetism (spontaneous circulating currents
around plaquettes); (ii) $p$-wave superconductivity 
(see the speculations on Fig.~\ref{fig:newhops}(b));
or (iii) the analog of spin-charge separation, the spinon being
replaced by a spinless particle that carries Fermi 
statistics~\cite{senthil}.  If it transpires that
such states are hard to stabilize without spin, 
that would shed additional light on the Hubbard model; contrariwise, 
if they {\it are} stabilized, they may be easier to 
study in the spinless case, 
free from any background of low-energy spin excitations.

A crude comparison may be made of the spinless-fermion
model with the infinite-$U$ Hubbard model in the dilute regime.
In that case, each fermion excludes one site (its own) 
from {\it half} of the other fermions, not counting the
exclusion built in by Fermi statistics.
In the present spinless model each fermion excludes 
{\it four} sites from {\it all} other fermions, so in a
sense the hole-rich metal phase is ``jammed'' eight times
more effectively than in the Hubbard case. 
We expect,  then, that kinetic-energy-stabilized
stripes are far more robust 
in the present model than in the large-$U$ Hubbard case. 
In fact they are practically marginal in the present model, 
so this encourages the opinion that stripes are {\it not} stable 
in the short-range Hubbard (or $t$--$J$) model.

%%% \acknowledgments
We acknowledge support by the National Science Foundation under
grants DMR-9981744 and PHY94-07194.
C. L. H. thanks R.~McKenzie, 
G.~Uhrig, 
D.~Scalapino, D.~Khomski, 
M.~Troyer, and G.~G.~Batrouni for helpful discussions.

%%%%%%%%%%%%%%%%% FIGURE 1 %%%%%%%%%%%%%%%%%%%%

%%%%%%%%%%%%%%%%% FIGURE 2 %%%%%%%%%%%%%%%%%%%%

%%%%%%%%%%%%%%%%% FIGURE 3 %%%%%%%%%%%%%%%%%%%%

\begin{thebibliography}{10}


\bibitem{Fa99} P. Fazekas, {\it Lecture Notes on Electron
Correlation and Magnetism} (World Scientific, Singapore, 1999).

\bibitem{Ko67} W. Kohn, Phys. Rev. Lett. 
19, 789 (1967).

\bibitem{Cu71} J. R. Cullen and E. Callen, Phys. Rev. Lett. 26, 236-8 (1971).

\bibitem{halfmetal} 
R.~A.~de Groot {\it et al},
Phys. Rev. Lett.  50, 2024 (1983).

%%%%%%%%%%%%%%%%%%%%
\bibitem{FN-halffilled}
The simulation of this model by
J. E. Gubernatis {\it et al}, 
Phys. Rev. B 
32, 103 (1985), 
was limited to the half-filled case, with finite $V$ and nonzero temperature.


\bibitem{Mu95} 
G. Murthy and R. Shankar, J. Phys. Condens. Matt. 
7, 9155 (1995).

\bibitem{uhrig} 
G. S. Uhrig and R. Vlaming, Phys. Rev. Lett. 
71, 271 (1993);
G. S. Uhrig and R. Vlaming, Physica B 
206 \& 207, 694 (1995).

\bibitem{emery}
(a) V.~J.~Emery, S.~A.~Kivelson, and H.~Q.~Lin, 
Phys. Rev. Lett. 64, 475 (1990);
(b) V.~J.~Emery and S.~A.~Kivelson, 
Physica C 209, 597 (1993).

\bibitem{Pre94}
P. Prelov\v{s}ek and I.~Sega, 
Phys. Rev. B 49, 15241 (1994).

\bibitem{zaanen}
 J. Zaanen, M.~L.~Horbach, and W.~van~Saarloos,
 Phys. Rev. B 53, 8671 (1996).

\bibitem{Na97}
C. Nayak and F. Wilczek, Phys. Rev. Lett. 
78, 2465 (1997).

\bibitem{Es98}
H. Eskes {\it et al},
Phys. Rev. B 58, 6963 (1998).
The lattice strings in their model
may become ``smooth'', with 
{\it nondivergent} fluctuations in (\ref{eq:stripefluc});
that might occur in our model
with Hamiltonian (\ref{eq:Hamspinless})
if $V/|t|<\infty$ or if (\ref{eq:newhops}) is added.

\bibitem{white} 
S.~R.~White and D. ~J.~Scalapino, 
Phys. Rev. Lett. 81, 3227 (1998).

\bibitem{pryadko}
L.~P.~Pryadko {\it et al},
Phys.~Rev.~B60, 7541 (1999).

\bibitem{newzaanen} 
For an argument on the form of $\phi(d)$, 
see J.~Zaanen 
Phys. Rev. Lett. 84, 753 (2000). 

\bibitem{Tr94} 
J. M. Tranquada {\it et al},
Phys. Rev. Lett. 73, 1003 (1994).

\bibitem{bosons-XXZ}
The hard-core boson model maps to a spin-1/2 $XXZ$ 
quantum antiferromagnet, in which occupied sites correspond to
up spins, 
the XY spin components have coupling $t$, and the $Z$ components
have coupling $V$. But this is not helpful
in our $V\to\infty $ limit, since the spin model 
Hamiltonian also contains a magnetic field of order $V$.

\bibitem{t-matrix} 
J.~Kanamori, Prog. Theor. Phys. 
30, 275 (1963);
V.~M.~Galitskii,
Sov. Phys. JETP 
34, 104 (1958).

\bibitem{mila}
F. Mila, Phys. Rev. B 49, 14047 (1994).
Mila actually considered a spinfull extended Hubbard model
but the spin freedom is trivial
in the $V=\infty$ case, since all $2 ^M$ assignments of fermion spins 
are degenerate eigenstates.

\bibitem{FN-walls}
Quantum-fluctuating strings in two dimensions, as
represented by a path integral, map
to {\it classical} fluctuating surfaces 
in three dimensions. The classical behaviors are reviewed
in, {\it e.g.}, M.~E.~Fisher, 
J. Stat. Phys. 
34, 667 (1984)


\bibitem{batrouni}
G.~G.~Batrouni {\it et al}
Phys. Rev. Lett. 74, 2527 (1995);
G.~G.~Batrouni  and R.~T.~Scalettar, 
Phys. Rev. Lett. 84, 1599 (2000). 

\bibitem{batrounipc}
A quantum Monte Carlo simulation at $T=1/6$ 
of the $4\times 16$ hardcore boson
system (with $V=\infty$), 
by G.~G.~Batrouni (personal communication), 
showed superfluid at $n_c \approx 0.23$, $\mu \approx 1.4$,  
coexisting with a dense (presumably 2-stripe)
state at $n=28/64$, consistent with
the phase diagram presented here, and
with our exact diagonalization result for the same $n$.
(This aspect ratio strongly favors stripe states.)
\bibitem{visscher} P.~Visscher, 
Phys.~Rev.~B 10, 943 (1974).
%%% Phase separation (1st mention for Hubbard)

\bibitem{FN-dilboson}
For bosons as $n\to 0$, 
one would expect $A_2 \to A_2 + A_2'/\ln n$ in
Eq.~(\ref{eq:Endilute}).
See M. Schick, Phys. Rev. A3, 1067 (1971). 

\bibitem{NG}
%%% These calculations are being extended.
N.~G.~Zhang and C.~L.~Henley, unpublished.

\bibitem{stripehole}
C.~L.~Henley, in preparation.

\bibitem{FN-stripetostripe}
$\Delta(L,L')$ depends slightly on $L'$ because holes can be transferred 
from one stripe to the next stripe.

\bibitem{wiese} 
S.~Chandrasekharan and U.-J.~Wiese, 
Phys. Rev. Lett.
83, 3116 (1999).

\bibitem{Noda99}
T.~Noda, H.~Eisaki, and S,~Uchida,  Science 286, 265 (1999).

\bibitem{senthil}
T.~Senthil and M.~P.~A.~Fisher, preprint (cond-mat/9910224),
and personal communications.

\end{thebibliography}
\end{document}